# Modulation doping to control the high-density electron gas at a polar/non-polar oxide interface

**Tyler A. Cain, Pouya Moetakef, Clayton A. Jackson, and Susanne Stemmer[a)]**

Materials Department, University of California, Santa Barbara, California, 93106-5050, USA.

**Abstract**

A modulation-doping approach to control the carrier density of the high-density electron gas at a prototype polar/non-polar oxide interface is presented. It is shown that the carrier density of the electron gas at a $GdTiO_3/SrTiO_3$ interface can be reduced by up to 20% from its maximum value (~ $3\times10^{14}$ $cm^{-2}$) by alloying the $GdTiO_3$ layer with Sr. The Seebeck coefficient of the two-dimensional electron gas increases concurrently with the decrease in its carrier density. The experimental results provide insight into the origin of charge carriers at oxide interfaces exhibiting a polar discontinuity.

[a] Electronic mail: stemmer@mrl.ucsb.edu

Interfaces such as Ge/GaAs [1,2], LaAlO$_3$/SrTiO$_3$ [3], and $R$TiO$_3$/SrTiO$_3$ [4-6], where $R$ is a trivalent rare earth ion, exhibit a polar discontinuity at the interface. An interfacial, high-density, mobile free electron (or hole) gas can serve to neutralize the divergent electrostatic potential energy created by the polar discontinuity. This mechanism is known as "electronic reconstruction". The origin of the free charge at LaAlO$_3$/SrTiO$_3$ interfaces has been a subject of significant debate in the literature [7-9]. One proposed explanation invokes electrons moving from the exposed polar surface of the LaAlO$_3$ to the interface [10,11]. Some support for this model is provided by the fact that mobile electrons are only observed when the LaAlO$_3$ layer exceeds a critical thickness [12]. This mechanism is similar to what is found in III-nitride heterostructures, where surface states are the source of free charge that compensates for built-in electric fields due to the discontinuity of a bulk polarization at the interface [13]. An exposed polar oxide surface has, however, a number of pathways available to solve its polar problem, including adsorbates and atomic reconstructions [14]. No dependence of the mobile carrier density on the thickness of the oxide layer with the exposed polar surface is expected in this case [15]. An alternative source of carriers, and mechanism to compensate for the polar problem, is *the interface itself*. Consider, for example, an atomically sharp $R$TiO$_3$/SrTiO$_3$ (001) interface, where the interface plane is TiO$_2$. The terminating $R^{+3}O^{-2}$ layer, which carries a +1 formal charge, can transfer ½ electron per interface unit cell, or ~ 3×10$^{14}$ cm$^{-2}$, to the interfacial TiO$_2$ plane, donating mobile charge to that layer. The mobile charge resides on one side of the interface (as determined by the band alignments) and will be attracted to the positive fixed charge at the interface, forming a confined two-dimensional electron gas (2DEG). For $R$TiO$_3$/SrTiO$_3$ interfaces, this description of interface charge appears to work well.  In particular, independent of the individual layer thicknesses and growth sequences, a confined [16,17] 2DEG is consistently observed at GdTiO$_3$/SrTiO$_3$ interfaces.  The 2DEG resides in the SrTiO$_3$, and has the required charge density, ~ 3×10$^{14}$ cm$^{-2}$ [6].

Further understanding of the origin of the mobile charge can be obtained by investigating the mobile charge density at polar/non-polar interfaces where the formal charges on the planes along the (001) surface normal deviate from the integer values of +1 and -1, respectively, in pure LaAlO$_3$ or $R$TiO$_3$.  In this Letter, we achieve this by alloying Gd$_{1-x}$Sr$_x$TiO$_3$ films and interfacing them with SrTiO$_3$.  Stoichiometric GdTiO$_3$ is a prototype Mott insulator [18], and alloying with Sr results in hole doping.  With increasing amount of Sr ($x$), the polar problem, as well as the charge on the Gd$_{1-x}$Sr$_x$O interface plane, are correspondingly reduced.  One thus expects a proportional reduction in the mobile charge density.  At $x \approx 0.20$, Gd$_{1-x}$Sr$_x$TiO$_3$ undergoes a insulator-to-metal transition [19], which sets an upper limit for controlling the density of the 2DEG with this approach.

Films were grown by molecular beam epitaxy (MBE) on insulating (001) (LaAlO$_3$)$_{0.3}$(Sr$_2$AlTaO$_6$)$_{0.7}$ (LSAT) substrates by co-deposition from elemental and metal-organic sources, respectively [20,21].  The Sr content in the Gd$_{1-x}$Sr$_x$TiO$_3$ films was controlled through the Sr flux by adjusting the temperature of the Sr effusion cell, between 325 and 375 °C.  X-ray photoelectron spectroscopy (XPS) was performed to estimate the Sr content.  Ohmic contacts were 50-nm-Ti/300 nm-Au contacts, where Au is the top layer.  Temperature dependent resistivity and Hall measurements were made using a physical properties measurement system (Quantum Design PPMS) in Van der Pauw geometry. Highly resistive $Gd_{1-x}Sr_xTiO_3$ films were measured in a rectangular geometry with a Keithley multimeter. In-plane Seebeck coefficient measurements [17] were made at room temperature. Two types of heterostructures were studied: $Gd_{1-x}Sr_xTiO_3$/LSAT and $Gd_{1-x}Sr_xTiO_3$/$SrTiO_3$/LSAT (see Table I).

Figure 1 shows the resistivity as a function of temperature for three $Gd_{1-x}Sr_xTiO_3$/LSAT samples with different $x$. While samples A1 ($x = 0$), A2 and A3 are insulating, the Sr content of A4 is sufficient for the film to be metallic (Fig. 1). The resistivity of the films decreases by more than three orders of magnitude (at room temperature) as $x$ is increased. Hall measurements of A4 showed $n$-type behavior with a carrier concentration of $7.4 \times 10^{21}$ cm$^{-3}$ at room temperature. The carrier concentration of metallic $R_{1-x}Sr_xTiO_3$ scales with $x$ [22,23], and allows for estimating $x$ to be about 0.44 for sample A4. Samples A1 - A3 were too resistive for Hall measurements. The Sr content of A2 and A3 was estimated by extrapolating from the composition of A4, and the vapor pressure of sublimated Sr [24] at the Sr cell temperature, assuming that Sr incorporates in proportion to the amount evaporated (which is a reasonable assumption, see ref. [21]). The compositions are listed in Table I. For samples A2 and A3, XPS measurements were in reasonable agreement with these estimates, i.e., $x = 0.05$ and 0.11 for A2 and A3, compared to values of $x = 0.04$ and 0.13 estimated from the growth conditions [25].

For the $Gd_{1-x}Sr_xTiO_3$/$SrTiO_3$/LSAT structures (B1-B4), the $Gd_{1-x}Sr_xTiO_3$ layers were grown under the conditions corresponding to those of A1-A4 (see Table I). Samples B1-B4 are metallic and $n$-type, even for $Gd_{1-x}Sr_xTiO_3$ layers where $x$ was below the insulator-to-metal transition, due to the formation of the high-density 2DEG at the interface. The sheet resistances [Fig. 2(a)] of heterostructures with insulating $Gd_{1-x}Sr_xTiO_3$ layers (B1-B3) exhibit a large decrease with decreasing temperature, which is characteristic for transport in the $SrTiO_3$, due to its increasing mobility with decreasing temperature [6]. In contrast, the sheet resistance of the sample with the metallic $Gd_{1-x}Sr_xTiO_3$ layer (B4) is relatively temperature independent, consistent with being dominated by the metallic film. The carrier concentration of sample B4 is $5.10\times10^{15}$ cm$^{-2}$ at room temperature, more than an order of magnitude greater than samples B1-B3. The Hall resistance is linear in magnetic fields up to 7 T at all temperatures for all samples, indicating that the Hall coefficient is dominated by the $n$-type 2DEG for samples B1-B3. The mobility of $Gd_{1-x}Sr_xTiO_3$ is very low (0.57 cm$^2$V$^{-1}$s$^{-1}$ at room temperature for sample A4), typical for Mott materials, so carriers in the $Gd_{1-x}Sr_xTiO_3$ have little influence on the measured Hall coefficient.

Most importantly, the sheet resistances [Fig. 2(a)] and the carrier densities [Fig. 2(b)] of the 2DEGs in the $Gd_{1-x}Sr_xTiO_3/SrTiO_3$ heterostructures with insulating $Gd_{1-x}Sr_xTiO_3$ (B1-B3) change in proportion with the Sr concentration in the $Gd_{1-x}Sr_xTiO_3$ layer. At room temperature, the sheet carrier density of B1 ($x = 0$) is $3.12\times10^{14}$ cm$^{-2}$, or approximately ½ electron per interface unit cell. The sheet carrier densities of the 2DEGs in B2 and B3 are $2.72\times10^{14}$ cm$^{-2}$ and $2.50\times10^{14}$ cm$^{-2}$, respectively, corresponding to a reduction in carrier density of 13% and 20% relative to sample B1. These results show that alloying of the $Gd_{1-x}Sr_xTiO_3$ films effectively modulates the density of the 2DEG. The low temperature (2 K) mobilities for these samples were between 279 and 425 cm$^2$V$^{-1}$s$^{-1}$, and sufficient to obtain Shubnikov-de Haas oscillations at high magnetic fields [26]. This modulation is also apparent in measurements of the thermopower (Seebeck coefficient), as discussed next.

Figure 3 shows the Seebeck coefficient of $Gd_{1-x}Sr_xTiO_3$/LSAT and $Gd_{1-x}Sr_xTiO_3$/SrTiO$_3$/LSAT heterostructures as a function of estimated $x$. The undoped GdTiO$_3$ film exhibits a positive (*p*-type) Seebeck coefficient of +153 µV/K, similar to bulk GdTiO$_3$ [27]. The Seebeck coefficient decreases with increasing Sr content to +145 µV/K and +69 µV/K, for samples A2 and A3, respectively. The decrease is consistent with an increasing hole concentration [28]. The metallic sample (A4) has a negative Seebeck coefficient (-22 µV/K). The change in sign at the insulator-to-metal transition is typical of the rare earth titanates [28].

Because the magnitude of the Seebeck coefficient decreases with increasing carrier concentration of the 2DEG [17], one expects more negative values of the Seebeck coefficient of the 2DEG in the $Gd_{1-x}Sr_xTiO_3$/SrTiO$_3$/LSAT structures with increasing Sr concentration $x$. As seen in Fig. 3 this is indeed the case. The Seebeck coefficient increases in value from -258 µV/K for samples B1 and B2 to -295 µV/K for sample B3. We note that the Seebeck coefficient of samples B1-B3 is dominated by the 2DEG, due to its low resistivity. Specifically, for two parallel-connected layers (i.e., the 2DEG and the $Gd_{1-x}Sr_xTiO_3$, which are denoted by subscripts *a* and *b*), the measured Seebeck coefficient (*S*) is given by [29]:

$$S = \frac{\frac{S_a}{R_{s,a}} + \frac{S_b}{R_{s,b}}}{\frac{1}{R_{s,a}} + \frac{1}{R_{s,b}}} \quad (1)$$

where $R_{s,i}$ is the sheet resistance of the layers. Because of the large difference in the sheet resistances (see Figs. 1 and 2), $S_a/R_{s,a}$ of the 2DEG is several orders of magnitude greater than $S_b/R_{s,b}$ of $Gd_{1-x}Sr_xTiO_3$. Thus, the $Gd_{1-x}Sr_xTiO_3$ layers do not contribute significantly to the measured $S$. The increase in magnitude of the Seebeck coefficient of sample B3 can therefore be attributed to the reduced 2DEG carrier density. The Seebeck coefficient of sample B4 is a composite of the metallic $Gd_{1-x}Sr_xTiO_3$ layer and the $SrTiO_3$ layer, and thus more complicated to interpret in terms of the individual contributions.

In summary, measurements of the carrier densities and Seebeck coefficients of $Gd_{1-x}Sr_xTiO_3/SrTiO_3$ heterostructures show that reducing the polar discontinuity and fixed charge at the interface through alloying can be used to modulate the mobile carrier density in a high-density electron gas that forms at oxide interfaces exhibiting a polar discontinuity. The results shed light on the origin of the charge carriers at this interface. As the fixed charge of the $(Gd_{1-x}Sr_xO)$ layer at the interface is reduced from its formal +1 value in pure GdO, the density of electrons donated to the *d*-band states of the interfacial $TiO_2$ layer is proportionally reduced. The experiments also show that the reduction in the 2DEG carrier density is somewhat greater than the estimated $x$, i.e., a 20% reduction in the 2DEG density for $x \approx 0.13$. The extra reduction may be attributed to other effects such as recombination with holes from the $Gd_{1-x}Sr_xTiO_3$ layer. Models of the mobile charge distribution and band bending obtained using a self consistent Poisson-Schrödinger solver [30] confirm that a modest additional reduction (relative to $x$) in the 2DEG density is expected. The results attest to the excellent control of charge densities that is possible with MBE grown layers. Finally, we note that the approach presented here can be considered as analogous to modulation doping, as developed for AlGaAs/GaAs interfaces [31]. Similar to AlGaAs/GaAs, the charge density in the 2DEG is modulated through doping of one of the layers comprising the interface. Here, the approach is used to reduce, rather than increase, the charge density.

The authors thank Tom Mates for the XPS measurements and Dan Ouellette for assistance with the resistivity measurements. We acknowledge support through the Center for Energy Efficient Materials, an Energy Frontier Research Center funded by the DOE (Award Number DESC0001009). T.A.C. also received support from the Department of Defense through a NDSEG fellowship. C.A.J. was supported by the MRSEC Program of the National Science Foundation under Award No. DMR-1121053, which also supported the MRL Central Facilities that were used in this study. This work made use of the UCSB Nanofabrication Facility, a part of the NSF-funded NNIN network

**References**

[1]   W. A. Harrison, E. A. Kraut, J. R. Waldrop, and R. W. Grant, Phys. Rev. B **18**, 4402 (1978).

[2]   H. Kroemer, J. Cryst. Growth **81**, 193 (1987).

[3]   A. Ohtomo and H. Y. Hwang, Nature **427**, 423 (2004).

[4]   S. Okamoto and A. J. Millis, Nature **428**, 630 (2004).

[5]   J. S. Kim, S. S. A. Seo, M. F. Chisholm, R. K. Kremer, H. U. Habermeier, B. Keimer, and H. N. Lee, Phys. Rev. B **82**, 201407 (2010).

[6]   P. Moetakef, T. A. Cain, D. G. Ouellette, J. Y. Zhang, D. O. Klenov, A. Janotti, C. G. Van de Walle, S. Rajan, S. J. Allen, and S. Stemmer, Appl. Phys. Lett. **99**, 232116 (2011).

[7]   S. A. Chambers, Surf. Sci. **605**, 1133 (2011).

[8]   D. G. Schlom and J. Mannhart, Nat. Mater. **10**, 168 (2011).

[9]   M. Huijben, A. Brinkman, G. Koster, G. Rijnders, H. Hilgenkamp, and D. H. A. Blank, Adv. Mater. **21**, 1665 (2009).

[10]  N. Nakagawa, H. Y. Hwang, and D. A. Muller, Nature Materials **5**, 204 (2006).

[11]  J. Mannhart and D. G. Schlom, Science **327**, 1607 (2010).

[12]  S. Thiel, G. Hammerl, A. Schmehl, C. W. Schneider, and J. Mannhart, Science **313**, 1942 (2006).

[13]  J. P. Ibbetson, P. T. Fini, K. D. Ness, S. P. DenBaars, J. S. Speck, and U. K. Mishra, Appl. Phys. Lett. **77**, 250 (2000).

[14]  C. Noguera, J. Phys.-Condes. Matter **12**, R367 (2000).

[15]   M. Stengel, Phys. Rev. Lett. **106**, 136803 (2011).

[16]   P. Moetakef, D. G. Ouellette, J. R. Williams, S. J. Allen, L. Balents, D. Goldhaber-Gordon, and S. Stemmer, submitted for publication (2012).

[17]   T. A. Cain, S. Lee, P. Moetakef, L. Balents, S. Stemmer, and S. J. Allen, Appl. Phys. Lett. **100**, 161601 (2012).

[18]   H. D. Zhou and J. B. Goodenough, J. Phys.: Condens. Matter **17**, 7395 (2005).

[19]   M. Heinrich, H. A. K. von Nidda, V. Fritsch, and A. Loidl, Phys. Rev. B **63**, 193103 (2001).

[20]   B. Jalan, R. Engel-Herbert, N. J. Wright, and S. Stemmer, J. Vac. Sci. Technol. A **27**, 461 (2009).

[21]   B. Jalan, P. Moetakef, and S. Stemmer, Appl. Phys. Lett. **95**, 032906 (2009).

[22]   Y. Tokura, Y. Taguchi, Y. Okada, Y. Fujishima, T. Arima, K. Kumagai, and Y. Iye, Phys. Rev. Lett. **70**, 2126 (1993).

[23]   T. Okuda, K. Nakanishi, S. Miyasaka, and Y. Tokura, Phys. Rev. B **63**, 113104 (2001).

[24]   M. W. Chase, J. L. Curnutt, A. T. Hu, H. Prophet, A. N. Syverud, and L. C. Walker, J. Phys. Chem. Ref. Data **3**, 311 (1974).

[25]   See supplemental material at [link to be inserted by publisher] for the XPS data and the modelled charge density.

[26]   P. Moetakef, D. G. Ouellette, J. R. Williams, S. J. Allen, L. Balents, D. Goldhaber-Gordon, and S. Stemmer, unpublished.

[27]   G. V. Bazuev and G. P. Shveikin, Inorg. Mater. **14**, 201 (1978).

[28]  C. C. Hays, J. S. Zhou, J. T. Markert, and J. B. Goodenough, Phys. Rev. B **60**, 10367 (1999).

[29]  P. Pichanusakorn and P. Bandaru, Mater. Sci. Eng. R **67**, 19 (2010).

[30]  M. Grundmann, BandEng program (unpublished).

[31]  R. Dingle, H. L. Stormer, A. C. Gossard, and W. Wiegmann, Appl. Phys. Lett. **33**, 665 (1978).

**Table I**

Table I: Summary of samples structures and estimated $x$.

| Sample number | A1 | A2 | A3 | A4 | B1 | B2 | B3 | B4 |
|---|---|---|---|---|---|---|---|---|
| $x$ (%) | 0 | 4 | 13 | 44 | 0 | 4 | 13 | 44 |
| SrTiO$_3$ thickness (nm) | - | - | - | - | 65 | 65 | 65 | 65 |
| Gd$_{1-x}$Sr$_x$TiO$_3$ thickness (nm) | 11 | 22 | 22 | 22 | 3 | 11 | 11 | 11 |

**Figure Captions**

**Figure 1**: Resistivity of $Gd_{1-x}Sr_xTiO_3$ thin films on LSAT as a function of temperature. For samples A1-A3, only every $20^{th}$ measured data point is shown for clarity.

**Figure 2**: Sheet resistance (a) and sheet carrier density (b) of $Gd_{1-x}Sr_xTiO_3/SrTiO_3/LSAT$ heterostructures as a function of temperature.

**Figure 3**: Room temperature Seebeck coefficient of $Gd_{1-x}Sr_xTiO_3/LSAT$ and $Gd_{1-x}Sr_xTiO_3/SrTiO_3/LSAT$ heterostructures as a function of estimated Sr concentration, $x$.

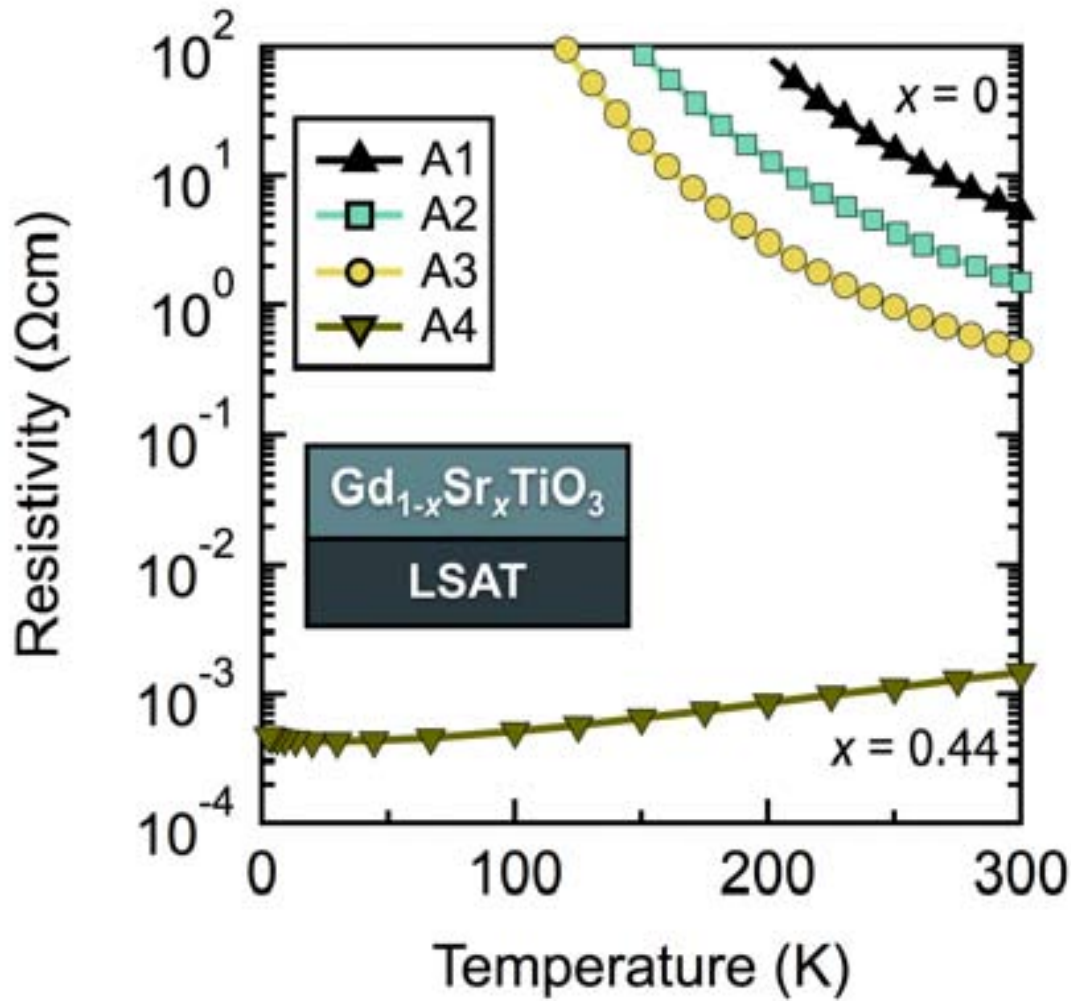

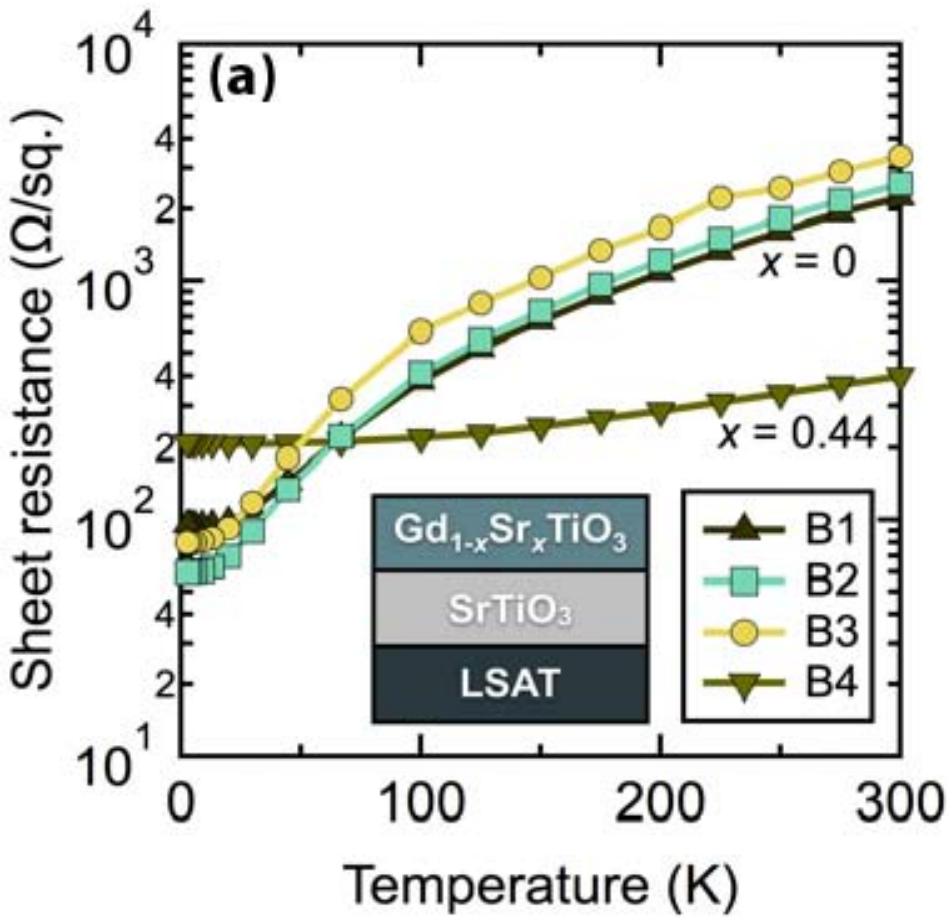

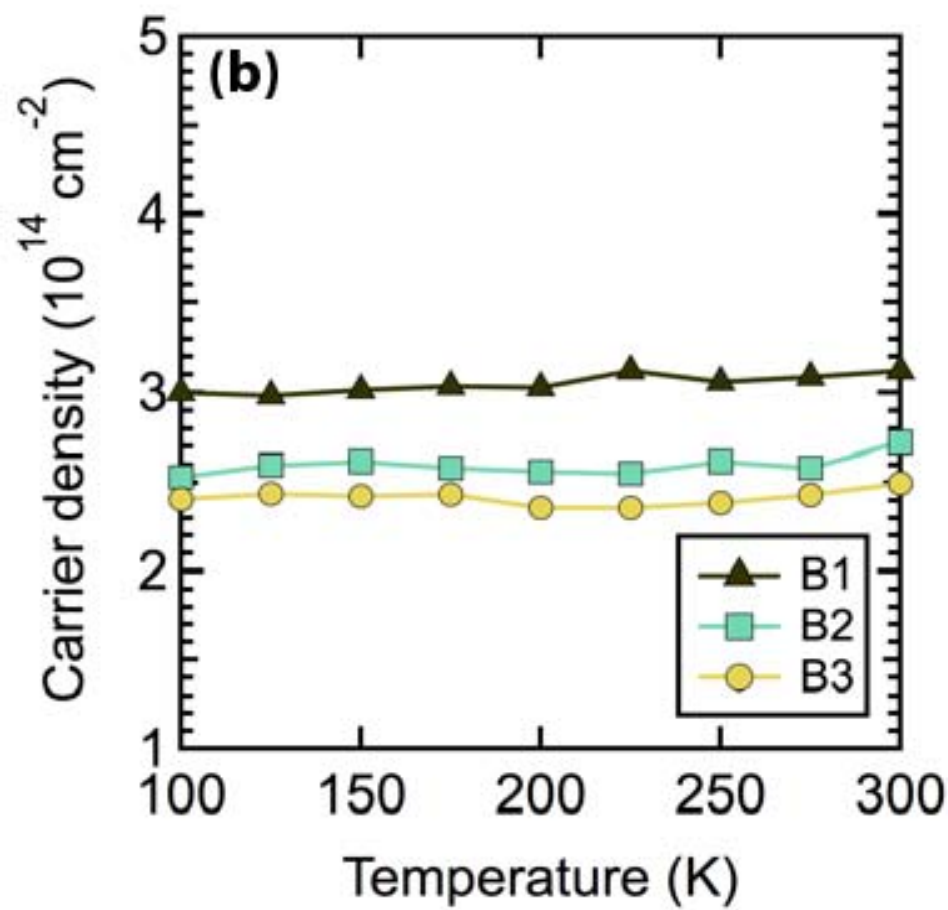

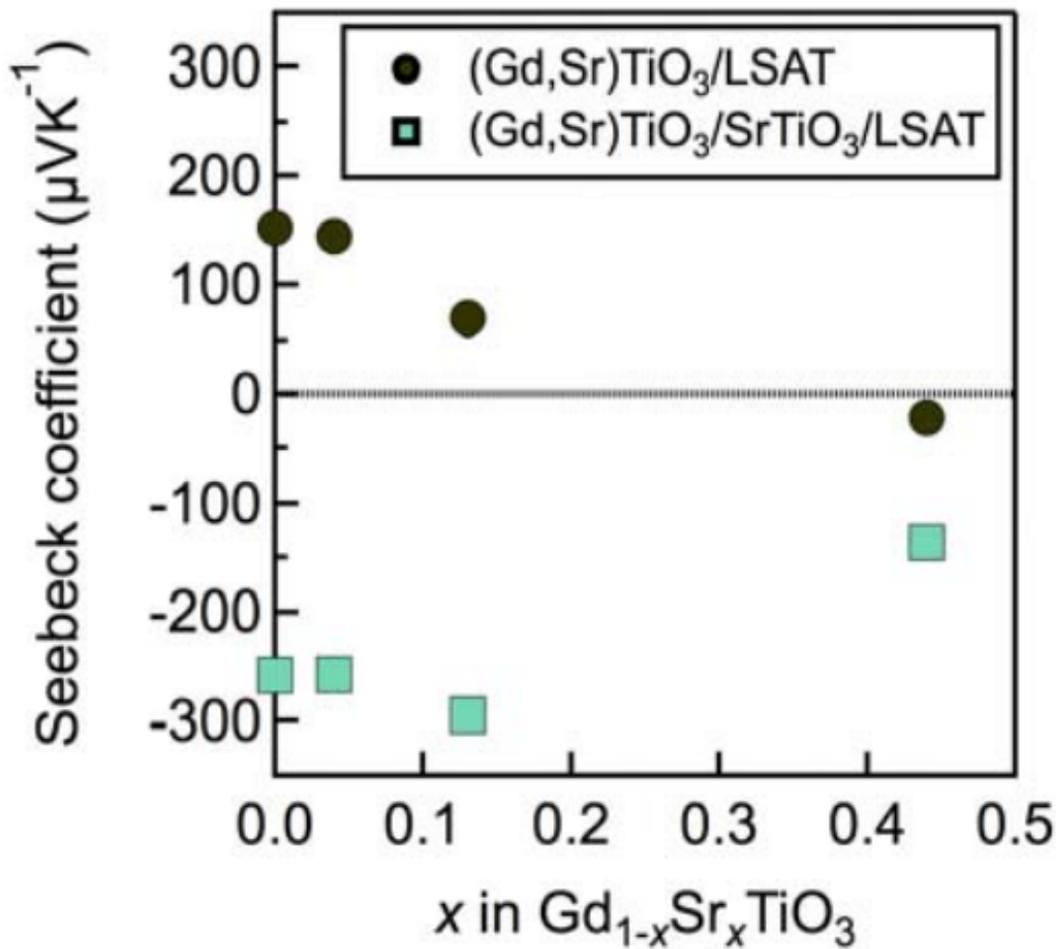